\documentclass[12pt]{iopart}
\usepackage{graphicx,cite}
\usepackage{bm}% bold math

\begin{document}
\title{Bifurcations, order, and chaos in the Bose-Einstein condensation
of dipolar gases}
\author{Patrick K\"oberle, Holger Cartarius, Toma\v z Fab\v ci\v c,
J\"org Main, G\"unter Wunner}
\address{Institut f\"ur Theoretische Physik 1, Universit\"at Stuttgart,
 70550 Stuttgart, Germany}
\date{\today}

\begin{abstract}
We apply a variational technique to solve the 
time-dependent Gross-Pitaevskii equation
for Bose-Einstein condensates
in which an  additional dipole-dipole interaction between the atoms is present
with the goal of modelling the dynamics of such condensates.
We show that universal stability thresholds for the collapse of the condensates
correspond to bifurcation points 
where always two stationary solutions of the Gross-Pitaevskii equation disappear in 
a tangent bifurcation, one dynamically stable and 
the other unstable. We point out that the thresholds also correspond
to  ``exceptional points'', i.e.\ branching singularities of the Hamiltonian.
We analyse the dynamics of excited condensate wave functions via Poincar{\'e}
surfaces of section for the condensate parameters and find both regular and chaotic motion, corresponding
to (quasi-) periodically oscillating and irregularly fluctuating condensates,
respectively.
Stable islands are found to persist up to energies well above the saddle
point of the mean-field energy, alongside with collapsing modes. The results
are applicable when  the shape of the condensate is axisymmetric.
\end{abstract}


%\maketitle

\bigskip\bigskip

At sufficiently low temperatures
a condensate of weakly interacting bosons can be represented by a single wave function 
whose dynamics obeys the dynamics of the Gross-Pitaevskii equation \cite{gross61,pitaevskii61}. The
equation is nonlinear in the wave function, and therefore the solutions of
the equation exhibit features not familiar from solutions of  ordinary Schr\"odinger equations of quantum mechanics.
As an example of the effects of the nonlinearity, Huepe et \textit{al}.\ \cite{hue99,hue03} demonstrated that for Bose-Einstein condensates
with attractive contact interaction, described by a negative  $s$-wave scattering length $a$, {\em bifurcations} 
of the stationary solutions of the Gross-Pitaevskii equation appear.
These authors also determined both the stable
(elliptic) and the unstable (hyperbolic) branches of the solutions. In physical terms, the bifurcation points correspond to critical
particle numbers, above which, for given strength of the attractive interaction, collapse of the condensate sets in.
For Bose-Einstein condensates of $^7$Li \cite{sacket99,gerton00}
and $^{85}$Rb atoms \cite{donley01,roberts01} these collapses were  experimentally observed.
 
In those condensates the short-range contact interaction is the only interaction to be considered. By contrast, in Bose-Einstein
condensates of {\em dipolar} gases \cite{santos00,baranov02,goral02a,goral02b,giovanazzi03} also a long-range dipole-dipole interaction is present.
This offers the unique
opportunity to study degenerate quantum gases with adjustable long-range {\em and} short-range interactions. The
achievement of Bose-Einstein condensation in a gas of chromium atoms  \cite{griesmaier05}, with a large dipole moment, has in fact opened the
way to promising experiments on dipolar gases  \cite{stuhler05}, which could show a wealth of novel phenomena \cite{giovanazzi02,santos03,li04,dell04}.
In particular,
the experimental observation of the collapse of dipolar quantum gases has also been reported  \cite{koch08} which occurs when 
the contact interaction is reduced, for a given particle number, below some critical value using a Feshbach resonance.

In this experimental situation it is most timely and appropriate to extend the investigations of the effects of the
nonlinearity of the Gross-Pitaevskii equation to dipolar quantum gases, and this is the goal of the present paper. 
To model these effects we will pursue, for the sake of simplicity, a variational ansatz.
We do this in the spirit of Refs.\ \cite{garcia97,hue99,hue03} where also variational techniques
were applied to model the dynamics of dilute ultracold atom clouds in the
Bose-Einstein condensed phase by solving the  Gross-Pitaevskii equation without dipole-dipole interaction.
In fact, quite recently Parker et al.\ \cite{Par08} have pointed out that in dipolar Bose-Einstein condensates
the Gaussian variational method gives excellent agreement with the full numerical solutions of the 
Gross-Pitaevskii equation in wide ranges of the physical parameters.
For oblate dipolar Bose-Einstein condensates the Gaussian approximation appears to agree only for weak
dipolar interactions.
The approximation used may not be valid in the limit of interest and clearly further exact studies based on
exact solutions of the dipolar Gross-Pitaevskii equation need to be carried out to verify the results. 
Full numerical quantum calculations, for condensates with only the contact interaction \cite{garcia97,hue99,hue03} present, 
or with an additional attractive gravity-like $1/r$ interaction \cite{ODe00,Pap07,Car08a,Car08b}, have confirmed 
that properties of the solutions of the Gross-Pitaevskii equation found in the variational calculations 
are recovered in the quantum calculations. We should therefore expect that a simple variational approach will also
capture essential features of the dynamics of condensate wave functions of dipolar gases.

We treat the problem in the ``atomic'' units provided by the magnetic
dipole-dipole interaction, i.e., we measure lengths in units of the
``dipole length''
\begin{equation}
\label{dipolelength}
 a_{\rm d} = \frac{\mu_0 \mu^2 m}{2\pi \hbar^2} = \frac{\alpha^2}{2}
 \frac{m}{m_e} \left(\frac{\mu}{\mu_{\rm B}} \right)^2 a_0 \, ,
\end{equation}   
energies in units of  $E_{\rm d} = \hbar^2/(2m a_{\rm d}^2)$, frequencies
in units of $\omega_{\rm d} = E_{\rm d}/\hbar$ and time in units of $\hbar/E_{\rm d}$, respectively.
In (\ref{dipolelength}), $\alpha$ is the fine-structure constant, $a_{0}$ the
Bohr radius, $m/m_e$ the ratio of the atom and electron mass, and $\mu$ the
magnetic moment.
For $^{52}$Cr, with $\mu = 6 \mu_{\rm B}$ ($\mu_{\rm B}$ the Bohr magneton),
one has $a_{\rm d} = 91~a_{0}$, $E_{\rm d} = 1.7\times 10^{-8}$~eV, and
$\omega_{\rm d}/2 \pi = 4.2 \times 10^{6}$~Hz.

In these ``atomic'' units, the Hartree equation of the ground state of a
system of $N$ identical bosons in an external trapping potential
$V_{\rm trap} = m( \omega_r^2 r^2 + \omega_z^2 z^2)/2$, all in the same
single-particle orbital $\psi$, interacting via the contact interaction
{\em and} the magnetic dipole-dipole
interaction, assumes the dimensionless form
\begin{eqnarray}
\label{HFat}
 &{\Big[}& \hskip -1mm - \Delta + \gamma_r^2 r^2 + \gamma_z^2 z^2
  +  N 8 \pi \frac{a}{a_d} |\psi({\bf r})|^2 \nonumber \\[1.1ex]
 &+&   N  \hskip -1mm  \int |\psi({{\bf r}\,}^\prime)|^2  \frac{1 - 3 \cos^2
  \vartheta^\prime}{|{{\bf r}}- {{\bf r}\,}^\prime |^3} d^3 {{\bf r}\,}^\prime
  \Big] \; \psi({\bf r}) = \varepsilon \, \psi({\bf r}). 
\end{eqnarray}
Here, $\varepsilon$ is the chemical potential, $a$ is the scattering length, and
$\gamma_{r,z} = \omega_{r,z}/(2 \omega_{\rm d})$ are the dimensionless
trap frequencies. Setting $\Psi = \sqrt{N}\psi$, one recovers the familiar form of the
time-independent Gross-Pitaevskii equation for dipolar quantum gases.

In the above ``atomic'' units,
the mean-field Hamiltonian in (\ref{HFat}) obeys a scaling
law with respect to the number $N$ of atoms:
Let $\tilde \psi(\tilde{\bf r})$ be a solution of the (formal) {\em one}-boson problem for
a given scattering length $a/a_{\rm d}$ and trap frequencies $\tilde\gamma_{r,z}$,
\begin{equation}
   H_{{\rm mf}}{(N=1, a/a_{\rm d}, \tilde \gamma_{r,z})}(\tilde{\bf r}) \; \tilde \psi(\tilde{\bf r})
 = \tilde \varepsilon \; \tilde \psi(\tilde{\bf r}) \; ,
\end{equation}
then {$ \psi ({\mathbf{r}} ):= \,N^{-3/2} \, \tilde \psi({\tilde{\bf r}})$}, with 
{${\bf r} = N{\tilde{\bf r}}$}, solves the {$N$}-bos\-on problem for the  
same scattering length {$a/a_{\rm d}$}, 
\begin{equation}\label{scaling}
H_{{\rm mf}}{(N, a/a_{\rm d},\gamma_{r,z})} \,({\bf r}) \; \psi\,({\bf r})  ={ \varepsilon} \; { \psi}\,({\bf r}) \,,
\end{equation}
but with trap frequencies $\gamma_{r,z}= \tilde \gamma_{r,z}/N^2$ and  chemical potential $\varepsilon = \tilde \varepsilon / N^{2}$. 
Note that the particle number scaling leaves the aspect ratio
$\lambda = \gamma_z/\gamma_r$ invariant.
Thus, the physical properties of Bose condensates of dipolar quantum gases quite generally
only depend on the value of the scattering length $a/a_{\rm d}$ and the 
particle number scaled trap frequencies $N^2 \gamma_{r,z}$ or, alternatively,
on the aspect ratio $\lambda$ and the scaled geometric mean of the trap
frequencies $N^2 {\bar{\gamma}} = N^2 \gamma_r^{2/3} \gamma_z^{1/3}$.
We note that the dimensionless parameter $D$ introduced by Dutta and Meystre \cite{dutta07}
and Ronen et \textit{al}.\ \cite{ronen07} to measure the effective strength of the dipole interaction
in trap frequency units is related to our scaling parameters by
$ D = (N^2 \gamma_r/2)^{1/2}$.

As an application of the particle number scaling law we emphasise that the experimental results
reported by Koch et \textit{al}.\ \cite{koch08} for the stabilisation of dipolar
chromium quantum gases for particle numbers $\sim$~20.000 and a trap frequency
$\bar{\omega}/2\pi = 720$~Hz correspond to a value of the scaled trap frequency 
$N^2 \bar{\gamma} = 3.4\times 10^{4}$. Therefore they  directly carry over to
any pairs of particle numbers and mean trap frequencies with this value of the
scaled trap frequency and the same aspect ratios!

To study the nonlinearity effects of the time-independent Gross-Pitaevskii equation for dipolar gases
we adopt the familiar variational ansatz of a (normalised) Gaussian type orbital 
(e.~g. \cite{santos00,li04,koch08})
\begin{equation}
\label{gaussian}
\psi({\bf r}) = A \exp \left[-(A_r r^2 + A_z z^2) \right]
\end{equation}
and exploit the time-dependent variational principle for the mean-field 
energy to determine the width parameters $A_r$ and $A_z$. It is well known that
solutions can be found only in certain parameter ranges 
and that at critical values
the condensate collapses. What -- to the best of our
knowledge -- for dipolar quantum gases has gone unnoticed before is that at these stability thresholds
actually {\em two} solutions of the extended Gross-Pitaevskii equation (\ref{HFat}) 
disappear.
\begin{figure}
\centering\includegraphics[width=0.7\columnwidth,clip=]{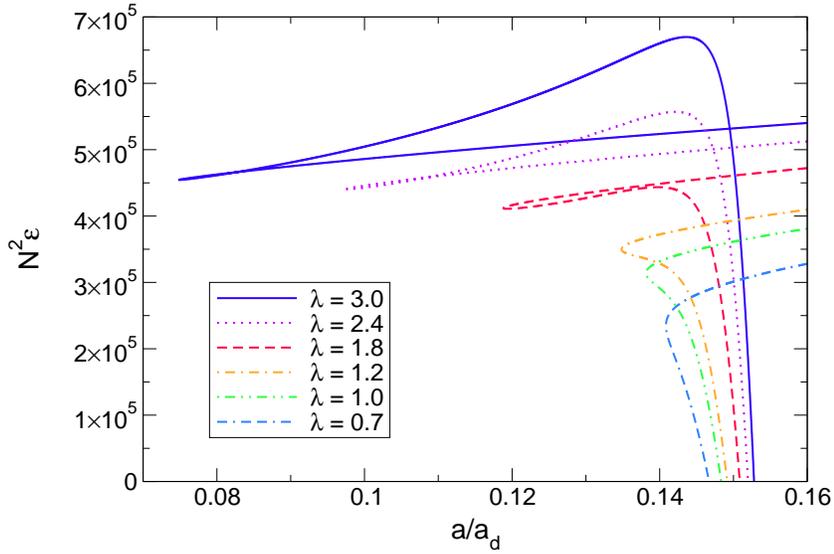}
\caption{\label{fig1}
  Chemical potential for the mean trap frequency ${N^{2} \bar{\gamma}} = 3.4\times 10^{4}$, used in the experiments of Koch et \textit{al}.\ \cite{koch08},
and different values of the aspect ratio. The tangential character of the bifurcations is particularly 
evident for the trap aspect ratios $\lambda \le 1.2$.}
\end{figure}
The situation is depicted in figure~\ref{fig1} where the chemical 
potential is plotted as a function of the scattering length for 
different values of the trap aspect ratio $\lambda$ 
and the fixed value of the scaled geometric mean of the 
trap frequency of $N^2 {\bar{\gamma}} = 3.4\times 10^{4}$. 
It can be seen that, as the scattering length is increased, for every value of $\lambda$ two solutions are born
in a tangent bifurcation, one corresponding to the ground state and the other
to a collectively excited state. 
An inspection of the mean-field energies shows that the excited
state corresponds to the branch of the chemical potential which diverges for $a/a_{\rm d} \to 1/6$.

\begin{figure}
\centering\includegraphics[width=0.7\columnwidth]{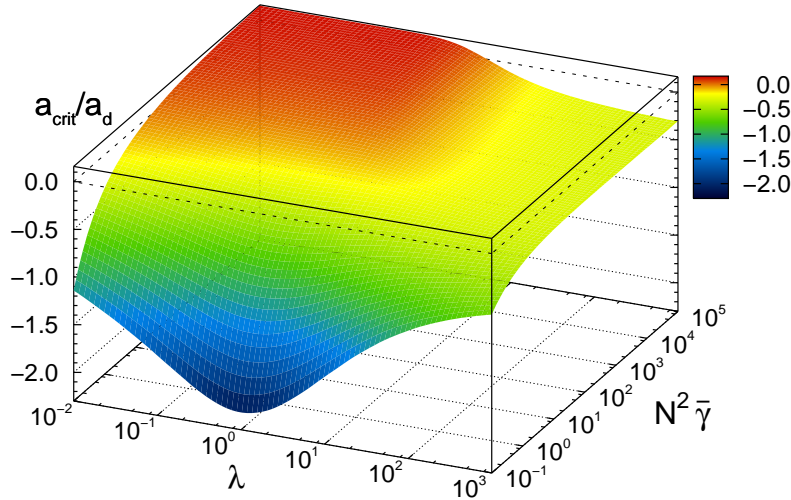}
\caption{\label{a_crit}
Stability thresholds, i.e.\ the critical scattering lengths $a_{\mathrm{crit}}/a_{d}$, as a function of the scaled trap parameter $N^{2} \bar{\gamma}$ 
and the aspect ratio $\lambda$. In the limit $N^{2} \bar{\gamma}
 \rightarrow \infty$, 
$\lambda \rightarrow 0$, one has $a_{\mathrm{crit}}/a_{d} = 1/6$.
}
\end{figure}

The behaviour of the stability thresholds over a wide region of the parameter space spanned by 
$N^{2} \bar{\gamma}$ and $\lambda$ is shown in figure~\ref{a_crit}. 
Large values of $N^2 \bar \gamma$ correspond to the regime where
the dipole-dipole interaction is dominant, as in the
experiments of Koch et \textit{al}.\ \cite{koch08}.  
For $N^2 \bar \gamma < 1$ the contact interaction prevails, and the dipole-dipole interaction is only
a small perturbation. 

In what follows we shall investigate the structure of the bifurcation,
the stability and the dynamics of dipolar condensates. 
The first aspect we want to highlight is 
that dipolar Bose-Einstein condensates near the
stability thresholds are experimental realizations of physical systems 
with ``exceptional points'',
a phenomenon hitherto observed only in open quantum systems, described by
non-Hermitian Hamiltonians (\cite{Kato66,Hei90}, see also
\cite{Car07b} and references therein). 
Exceptional points are positions in the parameter space where both the 
energy eigenvalues and the wave functions of (usually) two eigenstates pass 
through a branch point singularity as functions of the parameters and,
consequently, are identical at the critical set of parameters.
While in linear Schr\"odinger equations two eigenstates can only 
become identical if
the Hamiltonian is non-Hermitian, two coalescing eigenstates can also
occur for a nonlinear Schr\"odinger equation as an effect of its 
nonlinearity. 

This indeed the case for the stationary solutions of the Gross-Pitaevskii 
equation with dipole-dipole interaction.
At the stability thresholds the energy eigenvalues and the 
corresponding wave functions
of the two bifurcating states pass through a branch point singularity of the
Hamiltonian, the energies $N^{2}\varepsilon$ \emph{and} the
corresponding wave functions are identical. 
The way to reveal the branch point
singularity structure is to continue  the
scattering length $a/a_\mathrm{d}$ into the complex plane and to check a well-known 
property of exceptional points: If in traversing a full circle around the critical value
$a_{\mathrm{crit}}/a_\mathrm{d}$ at the point of bifurcation, 
$  a/a_\mathrm{d} = (a_{\rm crit}/a_\mathrm{d}) + \varrho\, \mathrm{e}^{i\varphi}, 
  \varphi = 0\dots 2\pi,
$
a permutation of the two solutions occurs, an exceptional point
is located within the circular area \cite{Kato66}.

\begin{figure}
  \centering\includegraphics[width = 0.7\columnwidth]{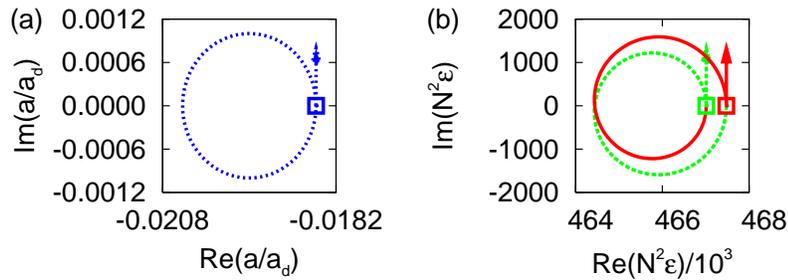}
  \caption{\label{fig:ep} Chemical potentials  $N^{2} \varepsilon$ (b) of
    the two stationary solutions emerging in the tangent bifurcation for a
    circle (a) with radius $\varrho = 10^{-3}$ around $a_{\mathrm{crit}}/a_\mathrm{d} = -0.01929$ in the complex extended 
parameter plane ($N^2 \bar{\gamma} = 3.4\times 10^4$,  $\lambda = 6.0$).
  The permutation of the chemical potentials after a full circle 
proves the existence of an exceptional point.}
\end{figure}

For the parameters of the Koch et \textit{al}.\ \cite{koch08} experiment 
we have performed
such an analysis, and for the case of an almost purely dipolar
quantum  gas the results are shown in figure\ \ref{fig:ep}. As one
goes around  a small circle  in the complex plane with the
critical scattering length at the centre, the two solutions  
\emph{are} permuted, i.e.\ we have an exceptional point.
It must be said, however,
that the usual way of experimentally proving the occurrence of an exceptional
point in open quantum systems, changing two real physical parameters
to traverse a circle in the complex energy plane, cannot be applied here.
In the case of the  Gross-Pitaevskii equation it is by continuing 
one real parameter, the scattering length, into the complex plane
that the nature of the stability thresholds as exceptional points is
revealed.

We now analyse the dynamics of the condensate wave functions. 
To do so we start from the time-dependent Gross-Pitaevskii equation,
which is (\ref{HFat}) with the replacement $\varepsilon\to i\frac{d}{dt}$,
and generalise the ansatz (\ref{gaussian}) to a time-dependent Gaussian
type orbital with complex width parameters $A_r(t)$ and $A_z(t)$ and complex normalisation
factor $A(t)$ (cf.\ \cite{garcia97}). We can then apply the time-dependent variational principle
 $\parallel i \dot{\psi}(t) - H \psi(t) \parallel ^2 {\buildrel ! \over =} \min$. \cite{Dir30,McL64}
to derive a system of ordinary nonlinear differential
equations for the time evolution of the real and imaginary parts of the
variational parameters $A_r(t)$ and $A_z(t)$. These can be shown  to be 
equivalent to canonical equations of motion belonging to a two-dimensional nonintegrable 
autonomous Hamiltonian system. For the time evolution one can therefore expect all the features familiar 
from nonlinear dynamics studies of such systems, including a transition to chaos.
For brevity, the  
details of these calculations are relegated to a subsequent paper. Here
we shall concentrate on the results.

We first investigate the stability of the two independent solutions
of the stationary Gross-Pitaevskii equation which emerge from the
bifurcations. To this end 
the four equations of motion for the real and imaginary parts of $A_r$ and
$A_z$ are linearised around the values corresponding to the stationary
states. In this way for each of the two states 
we obtain  four eigensolutions $\psi^\mathrm{(lin)} \sim e^{\kappa t}$ of the
linearised system of equations, with  eigenvalues $\kappa$.
For the ground state all eigenvalues turn out to be 
purely imaginary, proving  that the state is indeed dynamically stable.  
For the collectively excited state one also finds a positive real
eigenvalue, and hence the state is dynamically unstable. Similar
behaviour was found by Huepe et \textit{al}. \cite{hue99,hue03} in their
study of the stability of bifurcating solutions with only an attractive contact
interaction present.

The system of nonlinear 
first-order differential equations 
also serves to investigate the time evolution
of any initial state of the condensate by following the corresponding
trajectories in the four-dimensional configuration space spanned 
by the coordinates of the real and imaginary parts of $A_r$ and $A_z$.
Since the total mean-field energy is a constant of motion the
trajectories are restricted to three-dimensional hyperplanes, and their
behaviour can  most conveniently be visualised by two-dimensional 
Poincar{\' e} surfaces of section defined by requiring one of the 
coordinates to assume a fixed 
%real 
value. 

\begin{figure}
\centering\includegraphics[width=0.7\columnwidth,clip=]{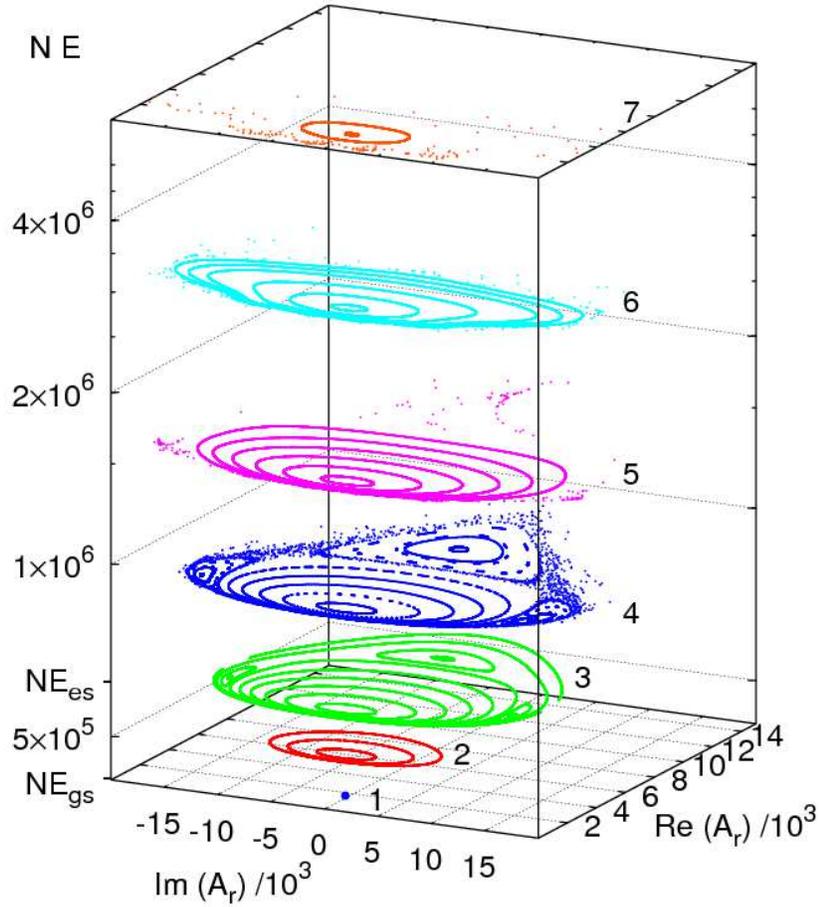}
\caption{Poincar\'e surfaces of section of the condensate wave functions represented by their width 
parameters for seven different mean-field energies at the scaled trap frequency $N^2 \bar{\gamma} = 3.4\times 10^{4}$, 
 aspect ratio $\lambda =6$, 
and the scattering length $a/a_{\rm d}=0.1$. The surfaces of section labelled 1 to 7 correspond to the increasing values
of the mean-field energy $NE = 4.24 \times 10^5, 5.00 \times 10^5, 6.00 \times 10^5, 9.00 \times 10^5, 1.50 \times 10^6, 3.00 \times 10^6,
6.00 \times 10^6 $, respectively. 
} 
\label{poincare}
\end{figure}

We  consider
Poincar{\' e} surfaces of section defined by the condition that 
the imaginary part of  $A_z(t)$ is zero. Each time the trajectory 
 crosses the plane ${\rm Im}{(A_z)} = 0$, the 
 real and  imaginary
parts of $A_r(t) = A_r^r(t) + i A_r^i(t)$ are recorded. 
In figure~\ref{poincare} surfaces of section  are plotted for 
seven different, increasing, values of
the mean-field energy. 
Again the physical parameters of the experiment of Koch
et \textit{al}.\ \cite{koch08}  are adopted, and the scattering length is fixed to $a/a_d=0.1$, away from its critical
value.
At these parameters, the variational mean-field energy of the ground state 
is $N E_{\rm gs} =4.24 \times 10^5$ and represents the local minimum on the two-dimensional
mean-field energy landscape, plotted as a function of the (real) width parameters. The variational energy
of the second, unstable, stationary state at these experimental parameters is $N E_{\rm es}=6.24 \times 10^5$, it corresponds to 
the saddle point on the mean-field energy surface. 
Between these two energy values the motion on the trajectories is 
bound, while for energies above the saddle point energy
the motion on the trajectories can become unbound: once the saddle point is traversed by 
a trajectory $A_r(t)$, $A_z(t)$, the parameters run to infinity, 
$A_r^r(t), A_z^r(t)\rightarrow \infty$, meaning a shrinking of the quantum state to vanishing width, i.e.\ a collapse of the condensate takes place.

The surface of section labelled 1 in figure~\ref{poincare} belongs to the energy of the ground state.
At this energy the kinematically allowed region for the crossing points of the trajectories is confined to a  single stable stationary point.
At the next higher energy (surface of section labelled 2) 
the kinematically accessible region in configuration space has grown, and
the initially stationary state has evolved into a periodic orbit (fixed point in the surface of section), corresponding to
a state of the condensate whose motion  is periodic. The oscillations of the width parameters
$A_r(t)$ and $A_z(t)$ represent oscillatory stretchings of the condensate along the $r$ and $z$
directions.  The  stable {\em periodic} 
orbit in the surface of section is surrounded by elliptical, quasi-periodic orbits, representing quasi-periodic
oscillations of the condensate. The surface of section 3, at the next higher energy, 
reveals that bifurcations have occurred, creating new stable and unstable periodic states, manifested
by the emergence of  additional elliptical islands and separatrices in the 
surface of section. 

The surface of section labelled 4  is the first in figure~\ref{poincare} with an energy value above the saddle point energy. 
Now chaotic orbits have appeared which surround the stable regions. 
In contrast to the (quasi-) periodic stretching oscillations 
of the condensate within the elliptical islands, the chaotic motion of the parameters describes
a condensate which does not yet collapse but whose widths  fluctuate irregularly.

One might imagine that well above the saddle point energy  stable condensate
wave functions no longer exist. However, in the surfaces of section 
labelled 5, 6, and 7 
regular islands are still clearly visible.
These stable islands are surrounded by chaotic trajectories. 
Since ergodic motion along  these trajectories comes 
close to every point in the configuration space, the chaotic motion sooner or later leads to a crossing of 
the saddle point and then to the collapse of the condensate wave functions.
It can be seen that with growing energy  
above the saddle point  the sizes of the stable regions gradually shrink. The reason why the kinematically allowed regions surrounding
the stable islands are hardly recognisable any more in these surfaces of section is that high above 
the saddle point energy the 
chaotic motion becomes more and more unbound, which means that the trajectories cross the Poincar{\' e} surfaces
of section only a few times, if ever, before they escape to infinity and  collapse takes place. 

It must be emphasised, however, that stable islands do persist even far above the saddle point energy, implying the existence of
quasi-periodically oscillating  nondecaying modes of the condensate wave functions.  

The prediction of such modes of energetically excited solutions of the
time-dependent Gross-Pitaevskii equation for cold dipolar
quantum gases is a result of our analysis. 
It would certainly be an intriguing and challenging task to examine whether it is 
possible to prepare excited states of dipolar
quantum gases of this type in both the regular as well as in the chaotic regions, to 
distinguish between the two different dynamics, and to access 
the stable regions high above the energy of the unstable stationary state. 
One way of creating the collectively excited states one might imagine is
to prepare the condensate in the ground state, and then to non-adiabatically reduce the trap frequencies.
Clearly experimental investigations along these lines are strongly encouraged.

One might ask, however,
whether the Gross-Pitaevskii equation underlying our calculations is adequate at all to describe complex dynamics 
of this type in real dipolar quantum gases. In particular in the chaotic regime local density maxima
might occur for which losses by two-body  or three-body collisions would have to be
taken into account. However, by virtue of the scaling law (\ref{scaling}) 
parameter ranges can always be found where the particle densities remain small even in 
these regimes and the Gross-Pitaevskii equation is applicable.

In this paper we have used a simple variational ansatz (\ref{gaussian}) to model 
the dynamics of axisymmetric solutions of the Gross-Pitaevskii
equation of dipolar gases. The advantage of the variational technique  is that
the analysis of the nonlinearity effects becomes especially transparent. 
Of course the results must be checked by accurate numerical simulations.
Such simulations, and the extensions to fully three-dimensional and 
structured condensate wave functions \cite{goral00,dutta07}, are under way.  
Comparisons of variational and accurate numerical results \cite{Pap07,Car08a,Car08b}
in an alternative
system with a long-range interaction, viz.\ Bose condensates with an attractive
gravity-like $1/r$ interaction \cite{ODe00}, have shown that the nonlinear dynamical properties found
in the variational calculation are recovered in the numerical calculations, and that the quantum
behaviour may be even richer. As a common feature of the  propagation of quantum wave functions,
the real-time evolution of arbitrary condensate 
wave functions of dipolar gases will exhibit complicated fluctuations. 
An interpretation of their full dynamics, and in particular
the search for periodic, quasiperiodic or chaotic structures, therefore will hardly be possible
without the guidance of the  variational results   obtained in this paper.

\ack
This work was supported by Deutsche Forschungsgemeinschaft. H.~C. 
is grateful for support from the Landesgraduiertenf\"orderung of
the Land Baden-W\"urttemberg.

\section*{References}

\bibliographystyle{unsrt}
%\bibliography{paper}

\end{document}